\documentclass{article}

\usepackage[dvips]{graphicx}
\usepackage{amsmath,amssymb}
\usepackage{algorithm}
\usepackage{algorithmic}
\usepackage{amsmath,amssymb,amsbsy,amsthm,setspace}

\let\epsilon\varepsilon
\let\phi\varphi

\let\epsilon\varepsilon

\newtheorem*{lemma*}{Lemma}
\newtheorem{theorem}{Theorem}

\newtheorem{corollary}{Corollary}
\newtheorem{claim}{Claim}
\newtheorem{definition}{Definition}

\begin{document}

\title{Two-faced processes and random number generators}

 \author{B. Ryabko$^{1,2}$\\ 
$^1$Institute of Computational Technology of Siberian Branch of \\  Russian Academy of Science,\\
$^2$Novosibirsk State University.
\\
  }

\date{}

\maketitle

\begin{abstract}
Random and pseudorandom number generators (RNG and PRNG) are
used for many purposes including cryptographic, modeling and simulation applications.
For
such applications a generated bit sequence should mimic true random, i.e., by definition, such a
sequence could be interpreted as the result of the flips of a “fair” coin with sides that are
labeled “0” and “1”. It is known that  the Shannon entropy of this process is 1 per letter, whereas 
for any other stationary process with binary alphabet the Shannon entropy is strictly less than 1.
On the other hand, the entropy of the PRNG output should be much less than 1 bit (per letter), but the output sequence
should look like truly random. 
We describe random processes for which those, in a first glance contradictory properties, are valid. 

More precisely, it is shown that there exist binary-alphabet random processes whose entropy is less than 1 bit (per letter), 
but a frequency of occurrences of any word $|u|$ goes to $2^{- |u|}$, where $|u|$ is the length of $u$.
In turn, it gives
a possibility to construct RNG and PRNG which possess theoretical  guarantees. 
 This, in turn,   is important for applications such as those in cryptography. 

\end{abstract}

\textbf{keywords:}  
Shannon entropy, random process, true randomness, random  number generator, pseudorandom number generator

\section{Introduction}

Random numbers are widely used in cryptographic, simulation (e.g., in  Monte Carlo methods)
and modeling  (e.g.,  computer games) applications. 
A generator of truly random binary digits generates such  sequences 
$x_1 x_2 ... $  that, with probability one, for any binary word $u$ the following property is valid:
\begin{equation}\label{fr}
  \lim_{t  \rightarrow\infty } \nu_t(u) / (t - |u| ) = 2^{-|u|} \, 
 \end{equation}
where $\nu_t(u) $ is a number of occurrences of the word $u$ in the sequence  $x_1 ... x_{|u|}$, 
$x_{2 } ... x_{ |u|+1 }, ...$, $x_{t-|u|+1} ... x_{t }$. 
(As in most studies in this field, for brevity, we will consider the case 
when processes generate letters fro the binary alphabet $\{0,1\}$,
 but the results can be extended to the case of any finite alphabet.) 
 The RNG and PRNG  attract attention of many researchers
due to its importance to practice 
and interest of theory, because, in a certain sense,
this  problem is close to foundations of probability theory, see, for example,
\cite{Calude:02,Vitanyi:08}. 

There are two  types of methods for generating sequences of random digits: so called RNG and PRNG.   
The RNGs are based on digitizing of physical processes (like noises in electrical circuits), whereas PRNGs 
can be considered as computer programs 
whose input is a (short) word (called a seed) and the output is 
 a long  sequence   (compared to the input). As a rule, the seed is a truly random sequence and the PRNG 
 can be viewed as an expander of randomness which stretches a short truly random  seed into a long sequence that is supposed
 to appear and behave as a true random sequence \cite{L'Ecuyer:15}. 
So, the purpose of RNG and PRNG is to use low-entropy sources for  generating sequences which look truly random. 
Note that the Shannon entropy of the truly random process (i.e., the Bernoulli with $p(0) = p(1) = 1/2$) 
 is 1 per letter, whereas
for any other stationary process the entropy is strictly less then 1; see \cite{Cover:06}. That is why, the properties
of truly randomness
and low entropy are, in a certain sense, contradictory.  

                There are a lot of papers devoted to RNG and PRNG, because they are widely used in 
cryptography and other fields. For example, the National Institute of Standards and Technology (NIST, USA)
published a recommendation specifying mechanisms for the generation of random bits using 
deterministic methods \cite{NIST-prng}. Nowadays, quality of almost all practically used RNG and PRNG is estimated by statistical tests
intended to find deviations from true 
 randomness (see, for ex., NIST  Statistical Test Suite \cite{NIST-test}). 
Nevertheless, researchers look for  RNG and PRNG with provable guarantees on their randomness 
 because methods with proven
properties are  of  great interest in cryptography. 

  In this paper we describe several kinds of random processes whose entropy can be much less than one, but, in a certain sense, 
  they generate sequences for which the property of true  randomness (\ref{fr}) 
 is valid either for any integer $k$ or 
  for $k$s from a certain interval (i.e. $1 < k < K$, where $K$ is an integer). 
  It shows the existence of  low-entropy RNGs and PRNGs which generate sequences satisfying the property (\ref{fr}).
  Besides, the description of the suggested processes show how they can be used to construct
  RNGs and PRNGs for which the property (\ref{fr}) is valid. 
 Note that so-called 
  two-faced processes, for which the property  (\ref{fr}) is valid for a given $k$ were described in \cite{BRyabko:05a,BRyabko:05}. 
 Here those processes are   generalized and  some new results concerning their properties  are established. 

More precisely, in this paper we describe the following two processes.
First, we describe  so-called two-faced process of order $k$, $k \ge 1$, which is the $k$-order Markov chain 
and, with probability 1,  for any  sequence $x_1 ... x_t$ 
and any binary word $u \in \{0,1\}^k $ the frequency of occurrence
of the word $u$ in the sequence $x_1 ... x_{|u|}$, 
$x_{2} ... x_{|u|+1 }, ...$, $x_{t-|u|+1} ... x_{t}$  goes to $2^{-|u|}$, where $t$ grows.  
Second, we describe  so-called twice two-faced processes for which this property is valid for any integer $k$.
Besides, we show how such processes can be used to construct  RNG and PRNG  for which the property (\ref{fr}) is valid.

The paper is organized as follows: the next part contains descriptions of two-faced processes and transformations. The  third part gives definitions  of the
so-called twice two-faced processes for which the property (\ref{fr}) valid for every 
 integer $k$.
In the conclusion we briefly discuss possible application of two-faced processes to RNG and PRNG.

\section{Two-faced processes}


First, we describe two families of random processes $T_{k, \pi}$ and
$\bar{T}_{k, \pi}$, where $k=1,2, \ldots,\,$ and $ \pi \in (0,1)$
are parameters. The processes $T_{k, \pi}$ and $\bar{T}_{k, \pi}$ are
Markov chains of the connectivity (memory) $k$, which generate
letters from $\{0,1\}$. It is convenient to define their transitional matrices
inductively. The process matrix of $T_{k, \pi}$ is defined by conditional
probabilities $P_{T_{1, \pi}}(0/0) = \pi, P_{T_{1, \pi}}(0/1) =
1-\pi $ (obviously, $P_{T_{1, \pi}}(1/0) =1- \pi, P_{T_{1, \pi}}(1/1) = \pi $). 
The process $\bar{T}_{1,\pi}$ is defined by
$P_{\bar{T}_{1, \pi}}(0/0) =1- \pi, P_{\bar{T}_{1, \pi}}(0/1) = \pi$. Assume that transitional matrices $T_{k, \pi}$ and $\bar{T}_{k, \pi}$ are defined and
describe $T_{k+1, \pi}$ and $\bar{T}_{k+1, \pi}$ as follows 
\begin{equation}\label{1}
P_{T_{k+1,  \pi}}(0/0u) = P_{T_{k, \pi}}(0/u), $$ $$ P_{T_{k+1, \pi}}(1/0u)
= P_{T(k, \pi)}(1/u), $$ $$ P_{T(k+1, \pi)}(0/1u) = P_{\bar{T}(k,
\pi)}(0/u), $$ $$ P_{T(k+1, \pi)}(1/1u) = P_{\bar{T}(k, \pi)}(1/u) ,
\end{equation}
and, vice versa, 
\begin{equation}\label{2} 
P_{\bar{T}(k+1, \pi)}(0/0u) = P_{\bar{T}(k,
\pi)}(0/u), $$ $$ P_{\bar{T}(k+1, \pi)}(1/0u) = P_{\bar{T}(k,
\pi)}(1/u), $$ $$ P_{\bar{T}(k+1, \pi)}(0/1u) = P_{T(k,
\pi)}(0/u), $$ $$ P_{\bar{T}(k+1, \pi)}(1/1u) = P_{T(k, \pi)}(1/u) 
\end{equation}
for each $u \in \{0,1\}^k$ (here $vu$ is a concatenation of the words
$v$ and $u$). For example,
\begin{equation}\label{ex}
  P_{T(2,\pi)}(0/00) = \pi,
P_{T(2,\pi)}(0/01) = 1-\pi, \end{equation}$$ P_{T(2,\pi)}(0/10) = 1-\pi,
P_{T(2,\pi)}(0/11) = \pi. $$

To define  a process  $x_1 x_2 ...$ the initial probability distribution needs to be specified. 
We define 
the initial distribution of the processes $T(k, \pi)$ and $\bar{T}(k, \pi)$,  $k=1,2, \ldots,\,$,
to be uniform on $\{0,1\}^k$, i.e. $P\{ x_1 ... x_k = u \} = 2^{-k}$ for any $u \in \{0,1\}^k$. 
On the other hand, sometimes processes with a
different (or unknown) initial distribution will be considered; that is why, 
in both cases the initial state will be mentioned in order to avoid
misunderstanding. 

Let us define the Shannon entropy of a stationary process $\mu$. The conditional entropy
of order $m$, $m=1,2, ...$, is defined by 
\begin{equation}\label{en}
 h_m = - \sum_{u \in \{0,1\}^{m-1} } \mu(u) \sum_{v \in \{0,1\} } \mu(v/u) \log \mu(v/u)
\end{equation} 
and the limit Shannon entropy is defined by  
\begin{equation}\label{enl}
 h_\infty  =  \lim_{m  \rightarrow \infty }h_m \, \, ,
\end{equation}
see \cite{Cover:06}.

 The following theorem describes the main properties of 
the   processes defined above. 

\begin{theorem}\label{T1}
Let a sequence $x_1 x_2 ... $ be 
generated by the process  $T(k, \pi)$ (or $\bar{T}(k, \pi)$), $k \ge 1$
and $u$ be a binary word of length $k$.
Then, 

i) If the initial state obeys the uniform distribution over $\{0,1\}^k$, then for any $j \ge 0$
 \begin{equation}\label{u2}  
 P(x_{j+1} ... x_{j+k} = u) = 2^{-|u|}  .
 \end{equation}  
 ii)  
 for any initial state of the Markov chain $T(k, \pi)$ (or $\bar{T}(k, \pi)$) 
\begin{equation}\label{u1}  
 \lim_{j  \rightarrow\infty } P(x_{j+1} ... x_{j+k} = u) = 2^{-|u|}  .
 \end{equation}  

 iii) For each $\pi \in (0,1) $ the $k$-order
Shannon entropy ($h_k$) of the processes $T(k, \pi)$ and
$\bar{T}(k, \pi)$ equals 1 bit per letter 
whereas the limit Shannon entropy ($h_\infty $) equals $ - (\pi
\log_2 \pi + (1- \pi) \log_2 (1-\pi) ).$ 
\end{theorem}

The proof of the theorem is given in the Appendix, but here we
consider examples of ``typical'' sequences of the processes
$T(1,\pi)$ and $\bar{T}(1,\pi)$ for $\pi$, say, 1/5. Such
sequences could be as follows: $ 010101101010100101...$ and $
000011111000111111000.... .$ We can see that each sequence
contains approximately one half of 1's and one half of 0's. (That
is why the first order Shannon entropy is 1 per a letter.) On the
other hand, both sequences do not look like truly random, because
they, obviously, have too long subwords like either  $101010 ..$
or $000.. 11111.. .$ (In other words, the second order Shannon
entropy is much less than 1 per letter.)  So,
informally, we can say that those sequences mimic truly random, if one takes into account only frequencies of words of the 
length one. 

Due to Theorem 1, we give the following 
\begin{definition}
 A random process is called asymptotically two-faced of order $k$,  if the equation (\ref{u1}) is valid for all 
 $u \in \{0,1\}^k$.
 If the equation (\ref{u2}) is valid, the process is called 
 two-faced of order $k$.
\end{definition}

Theorem 1 shows that the processes $T(k, \pi)$ and $\bar{T}(k, \pi)$ are two-faced. 
 The statements i) and ii) show that the processes look like truly random
if we consider blocks whose length is less than the process order $k$. On the other hand, if we take 
into consideration blocks whose  length is grater, the statement iii) shows that their distribution is far form uniform
(if $\pi$ is either  small or large). Those properties  explain the name ``two-faced''.

The following theorem shows that, in a certain sense, there exist quite many two-faced processes. 
\begin{theorem}\label{T2}
 Let $X = x_1 x_2 ... $  and $Y = y_1 y_2 ...$  be random processes. 
 We define the process $Z = z_1z_2 ...$ by equations $z_1 = x_1 \oplus y_1$, $z_2 = x_2 \oplus y_2, ...$ where $ x_1 x_2 ... $ and 
 $y_1 y_2 ...$ are 
 distributed according to $X$ and $Y$ and $ a \oplus b = ( a+b) \, mod \, \, 2 $. 
 Then, if $X$ is a $k$-order two-faced process ($k \ge 1$), then  $Z$ is a $k$-order two-faced process.  
  If $X $ is an asymptotically $k$-order two-faced process then $Z$ is asymptotically $k$-order two-faced, too.
\end{theorem}

\section{Two-faced transformation}
Earlier we described two-faced processes which, in a certain sense, mimic truly random. 
In this section  we show how 
any Bernoulli   process can be  converted  to a two-faced process.
Informally, any sequence $X = x_1 x_2 ... $ created by Bernoulli process with $P(x_i = 0) = \pi $, $P(x_i = 1) = 1 - \pi $, 
will be transformed into   a sequence $y_1 y_2 ... $ of ``letters'' $\pi$ and $(1-\pi)$ by a map $0 \rightarrow \pi$, 
$1 \rightarrow (1-\pi)$. Then this sequence can be considered as an  input of the transition matrix $T_{k,\pi} $ and a new 
sequence $Y= y_1 y_2 ... $ can be generated according to $k$-order two-faced process, if we have an initial state, i.e. a binary word of 
length $k$. 
For example, let $k = 2$, the initial state be $01$ and $x_1 x_2 ... x_5$ $= 10010$. Then, $y_1 ... y_5$ $= (1-\pi) \pi \pi (1-\pi) \pi$
and, according to (\ref{ex}), we obtain a new  sequence $01110$.
In fact, the output  sequence is generated by the transition matrix  $T_{k,\pi} $; that is why the output process
is $2$-order two-faced.

Now we formally describe  a family of transformations which, in a certain sense, convert   random  processes into two-faced ones.
For this purpose we first define two families of matrices $M_k$ and $\bar{M}_k$, $k \ge 1$, which are 
 connected 
 with transition matrices $T_{k,\pi} $ and $\bar{T}_{k, \pi}$.
\begin{definition} For any $k \ge 1$, $v \in \{0,1\}^k$,
$w \in \{0,1\}$, the matrix $M_k$ is defined as follows:
\begin{equation}\label{tau} 
M_k(w,v) = \left\{ \begin{array}{c}   \, 
0,\, \, \,  if \, \, \,  T_{k,\pi}(w,v) = \pi \\
\, \, \,  \, \, \,  1,  \, \, \, if  \, \, \, T_{k,\pi}(w,v) =  1 - \pi \\
 \end{array} \right. 
                                                      \end{equation}
 $\bar{M}_k$ is obtained from $\bar{T}_{k,\pi}$ analogously. 
\end{definition}
Informally, these matrices combine the two steps from the previous example. 
 Namely, 
a transition from $x_1 x_2 ... $ to a sequence of   symbols $\pi$, $1-\pi$ and,
second, transition from it to the new  sequence of zeros and ones.
\begin{definition}
 Let $X=  x_1 x_2 ... $ be an infinite binary word,  $k > 0$ be an integer and $v \in \{0,1\}^k$.  
 The two-faced conversion  $\tau^k$ maps a pair $(X,v)$ into an infinite binary sequence $Y$ as follows:
 $$ y_{-k+1} y_{-k+2} ... y_0 = v \, \, ,$$  
 \begin{equation}\label{tau}   
  y_i =    M_{k} ( x_i,  \, \, \,  y_{i-k} y_{i-k+1} \, ... \, y_{i-1} )\,  \, \,  \, \,   if  \, \, 1  \le i                                                       
                                                      \end{equation}
where $i = 1, 2, ... $.
\end{definition}
It can be seen from definitions that the $y_1 y_2 ... $ is generated according to the transition matrix $T_{k,\pi}$ if 
$x_1 x_2 .. $ generated by Bernoulli process with $P(0) = \pi$, $P(1) = (1-\pi)$. 
From this and Theorem 1 we obtain the following statement: 

\begin{claim} 
 Let $X =  x_1 x_2 ... $ be any Bernoulli process, $k \ge 1$ be an integer
 and $\tau_k $ be a two-faced transformation. If  $v$ is a word from $ \{0,1\}^k$,
 then $ \tau^k(X,v )$ is asymptotically  two-faced of order $k$. If, additionally, $v$
 obeys the uniform distribution on $\{0,1\}^k$, then $ \tau^k(X,v)$ is  two-faced of order $k$. 
\end{claim}

 \section{Generalization}

The $k$-order two-faced processes mimic true random ones for block lengths $1, 2, ..., k$. Here we describe
such processes that  mimic true randomness for blocks of every 
 length.
By analogy with  so-called twice universal codes known in  information theory, we call such processes  twice two-faced. 
\begin{definition}
 A random process is called (asymptotically) twice two-faced,  if the equation (\ref{u2}) ( (\ref{u1}) ) 
 is valid for every integer $k$ and $u \in \{0,1\}^k$. 
 \end{definition}
Now we describe a family of such processes.

Let $n^* = n_1, n_2, .... $ be an infinite sequence of integers such that $n_1 < n_2 < n_3 ....$ and 
$X^1 = x^1_1 x^1_2 ...$, $X^2 = x^2_1 x^2_2 ...$, $X^3 = x^3_1 x^3_2 ...$, $...$
be (asymptotically) two-faced
processes of order $n_1, n_2, ...$, correspondingly.  Define a process $W$ $= w_1 w_2 $ $ ...$ by
\begin{equation}\label{3} 
w_i =
 \left\{ \begin{array}{c}  
  x^1_i\ \ \,\, \,  \,\, \, \,\, \, \,\, \,\,  i\le n_1, \\ 
  x^1_i \oplus x^ 2_i \, \, \, \,\, \,  \,\, \, \,\, \,\,\, \, n_1 < i \le n_2 , \\
  x^1_i \oplus x^ 2_i \oplus x^3_i  \, \, \, \,   n_2 < i \le n_3 , \\
  ............................. \\
   \end{array} \right. 
   \end{equation}
   and denote this process as $\bigoplus_{i=1}^\infty X^i$.
\begin{theorem}\label{T3}   
If all $ X^i$ $i=1,2, ... $ are two-faced then the process  $\bigoplus_{i=1}^\infty X^i$ is
 twice two-faced, i.e. for any binary word $u$ the equation (\ref{u2}) is valid. If all 
$X^i$ are asymptotically two faced, 
then  the process  $\bigoplus_{i=1}^\infty X^i$ is asymptotically 
 twice two-faced, i.e. 
 equation (\ref{u1}) is valid for any word $u$.
\end{theorem}
\begin{corollary} 
 Let $X = x_1 x_2 ... $  and $Y = y_1 y_2 ...$  be random processes and $X$ be twice two-faced. 
 Then the process $Z$ defined 
by equations $z_1 = x_1 \oplus y_1$, $z_2 = x_2 \oplus y_2, ...$ is twice two-faced, too.
\end{corollary}

It is worth noting that the total entropy of the processes $X^1, X^2, ... $ can be arbitrarily small, hence,
the ``input randomness'' of the process $\bigoplus_{i=1}^\infty X^i$ can be very small, whereas, in a certain sense,
the process looks like  a true random one.

\section{Conclusion}

In this paper we  focus on the existence of processes  whose entropy can be arbitrary small, but they 
mimic truly randomness in the sense that the frequency of occurrences of  any
word $u$ asymptotically equals $2^{-|u|}$. In Conclusion we note how   such processes can be directly used in order to
construct (or ``improve '' ) RNGs and PRNGs. For example, Theorems 2 and 3 shows that output sequence of any RNG
and PRNG will, in a certain sense, looks like truly random, if it is summed 
 with a (low-entropy) two-faced (or twice two-faced) process.

The possibility to transform  Bernoulli processes into two-faced ones gives a possibility to create  low-entropy
two-faced processes. Indeed, schematically, it can be done as follows:
Imagine, that one has a short word $v$ (it corresponds to the seed of a PRNG) and wants to create a 
sequence $V$, $|V| > |v|$, which could be considered
as  generated by a $k$-order two-faced process.  Now denote $h= |v|/|V|$ and let $\pi$ be a solution of the equation
$- (\pi \log \pi + (1- \pi) \log (1-\pi) = h$. It is well-known in information theory  that
there exists a lossless
code $\phi$ which compresses sequences  generated by a Bernoulli process with probability $(\pi, 1 - \pi)$ in such a way that 
the (average) length of output words is close to the Shannon entropy $h$, see \cite{Cover:06}.
Denote the decoder by $\phi^{-1}$  and let the sequence $U $ be $ \phi^{-1}(v)$. Informally, this sequence will look like  
generated by a Bernoulli source with probabilities $(\pi, 1 - \pi)$ and the final sequence $V$ can be obtained from $U$ by the transformation as  described 
in the Definition 3. 
 (We did not consider the initial $k$-bit words, which can be obtained, for example, as a part of the seed $v$.
In such a case $h$ can be defined as $(|v|-k)/|V|$.

\section{Appendix}
\emph{Proof} of Theorem 1. We prove the theorem for the process
$T_{k, \pi},$  but this proof is valid for $\bar{T}_{k, \pi},$ too.
 First we show that
\begin{equation}\label{a} p^*(x_1...x_k)=2^{-k}, \end{equation}
 $ (x_1...x_{k}) \in \{ 0,1 \}^{k}, $ is a
stationary (or limit) distribution for the processes $T_{k, \pi}$. 
 For any values of $k, k \geq 1,$ (\ref{a}) will be proved if we
show that the system of equations
$$ P_{T(k, \pi)}(x_1...x_k)=
P_{T(k, \pi)}(0x_1...x_{k-1})\,P_{T(k,\pi)}(x_k/0x_1...x_{k-1})\:
$$ $$+\,P_{T(k, \pi)}(1x_1...x_{k-1})\,
P_{T(k, \pi)}(x_k/1x_1...x_{k-1}) \, ;
$$ 
$$ \sum_{v \in \{0,1\}^k} p(v) = 1 $$
has the solution
$p(x_1...x_k)=2^{-k}$, $ (x_1...x_{k}) \in \{ 0,1 \}^{k}.$ It can be easily seen,  if we take
into account that, by definitions   (\ref{1}) and   (\ref{2}), the equality $P_{T(k, \pi)}(x_k/0x_1...x_{k-1})\: +\,
P_{T(k, \pi)}(x_k/1x_1...x_{k-1})=1 $ is valid for all $
(x_1...x_{k}) \in \{ 0,1 \}^{k} $. From this equality and the low
of total probability we immediately obtain (\ref{a}).
Having taken into account that the initial distribution matches the stationary (limit) one, 
we obtain the the first claim  of the theorem (\ref{u2}).
 From definitions (\ref{1}), (\ref{2}), we can see that all transition probabilities are nonzero 
 (they are either $\pi$ or $1 - \pi$). Hence, the Markov chain  $T(k, \pi)$ is ergodic and 
   the equations (\ref{u2}) are valid due to ergodicity. 

Let us prove the third claim of the theorem. From the definitions (\ref{1}), (\ref{2})
  we can see that either $P_{T(k,
\pi)}(0/x_1...x_{k})= \pi,\, P_{T(k, \pi)}(1/x_1...x_{k})=1-\pi$
or $P_{T(k, \pi)}(0/x_1...x_{k})=1- \pi,\, P_{T(k,
\pi)}(1/x_1...x_{k})=\,\pi$. From this and (\ref{en}) we can see that $h_{k+1} =
- (\pi \log_2 \pi + (1- \pi) \log_2 (1-\pi) )$ and, taking into account  (\ref{enl}),  we obtain  
$h_\infty = - (\pi \log_2 \pi + (1- \pi) \log_2 (1-\pi) )$. The
theorem is proved.

\emph{Proof} of Theorem 2.  The following chain of equations proves the first claim of the theorem:
\begin{equation}\label{f}
P\{z_{j+1} ... z_{j+k} = u\} = $$ $$\sum_{v \in \{0,1\}^k} P\{x_{j+1} ... x_{j+k} = v \} P\{y_{j+1} ... y_{j+k} = v\oplus u \} 
\end{equation}
$$ = 2^{-k} \sum_{v \in \{0,1\}^k}  P\{y_{j+1} ... y_{j+k} = u\oplus v \} =
2^{-k} \times 1 = 2^{-k} \, . $$
(Here we took into account (\ref{u2}) and the obvious equation $v\oplus u \oplus v = u$.)
In order to prove the second statement, we note that by definitions,  
$$
 \lim_{j  \rightarrow\infty } P(x_{j+1} ... x_{j+k} = u) = 2^{-|u|}  $$
 for any $u \in \{0,1\}^k$, see (\ref{u1}). Hence, for any $\delta$, $\delta > 0$,
 there exists $J$ such that 
 $$ |P(x_{j+1} ... x_{j+k} = u) - 2^{-|u|} | < \delta \, \, u \in \{0,1\}^k \, 
 $$ if $j > J$.
 From this inequality and the equation   (\ref{f}) we obtain 
 $$ (2^{-k}- \delta) \sum_{v \in \{0,1\}^k}  P\{y_{j+1} ... y_{j+k} = u\oplus v \} $$ $$
\le P\{z_{j+1} ... z_{j+k} = u\} \le $$ $$
(2^{-k} + \delta) \sum_{v \in \{0,1\}^k}  P\{y_{j+1} ... y_{j+k} = u\oplus v \} \, . $$
Taking into account that this sum equals 1, we obtain the following inequalities:
$$ (2^{-k}- \delta) 
\le P\{z_{j+1} ... z_{j+k} = u\} \le  
(2^{-k} + \delta)  \, . $$
It is true for any $\delta > 0$, hence (\ref{u2})
is valid and the process $Z$ is asymptotically $k$-order two-faced.
Theorem is proven.

\emph{Proof} of Theorem 3.  Let $ u$ be any binary word and $|u| = k $.
Take such an integer $n_i$ that $ k \le n_i$ and consider
the process $S = \bigoplus_{j=1}^{i-1} X^j \, \oplus \, \bigoplus_{j= i+1}^{\infty} X^j $.
(Here $U \oplus V = \{ u_1 \oplus v_1 \, \, u_2 \oplus v_2  \, \, u_3 \oplus v_3  \, \,... \, \}$ .
Obviously, $ \bigoplus_{j = 1}^{\infty} X^j =  X^i \oplus S$. The process $X^i$ is (asymptotically) $n_i$-order 
two faced. Having taken into account Theorem 2 we can see that $ \bigoplus_{j = 1}^{\infty} X^j $ is 
$n_i$-order two faced and, hence, $k$-order (asymptotically) two-faced (because $k \le n_i$, 
hence (\ref{u2}) (  (\ref{u1}) ) is valid.  
Theorem is proven.

\section*{Acknowledgment}
Research  was supported  by  Russian Foundation for Basic Research
(grant no. 15-29-07932).


\begin{thebibliography}{1}

\bibitem{NIST-prng}
E.~Barker and J.~Kelsey.
\newblock {\em Recommendation for Random Bit Generator (RBG) Constructions
  (DRAFT NIST Special Publication 800-90C)}.
\newblock National Institute of Standards and Technology, 2012.

\bibitem{Calude:02}
C.~S. Calude.
\newblock {\em Information and Randomness - An Algorithmic Perspective. 2nd
  Edition}.
\newblock Springer-Verlag, 2002.

\bibitem{Cover:06}
Thomas~M. Cover and Joy~A. Thomas.
\newblock {\em Elements of information theory}.
\newblock Wiley-Interscience, New York, NY, USA, 2006.

\bibitem{L'Ecuyer:15}
Pierre L'Ecuyer.
\newblock {\em Random Number Generation and Quasi-Monte Carlo}.
\newblock Wiley, 2014.

\bibitem{Vitanyi:08}
M.~Li and P.M.B. Vitanyi.
\newblock {\em An Introduction to {K}olmogorov Complexity and Its Applications,
  3rd edition}.
\newblock Springer, New York, 2008.

\bibitem{NIST-test}
Andrew Rukhin, Juan Soto, James Nechvatal, Miles Smid, Elaine Barker, Stefan
  Leigh, Mark Levenson, Mark Vangel, David Banks, Alan Heckert, James Dray, and
  San Vo.
\newblock {\em A Statistical Test Suite for Random and Pseudorandom Number
  Generators for Cryptographic Applications}.
\newblock National Institute of Standards and Technology, 2010.

\bibitem{BRyabko:05}
B.~Ryabko and A.~Fionov.
\newblock {\em Basics of Contemporary Cryptography for IT Practitioners}.
\newblock World Scientific Publishing Co, 2005.

\bibitem{BRyabko:05a}
B.~Ryabko and V.~Monarev.
\newblock Using information theory approach to randomness testing.
\newblock {\em Journal of Statistical Planning and Inference}, 133(1):95--110,
  2005.

\end{thebibliography}

\end{document}